\documentclass[aps,epsfig,preprint]{revtex4-1}
\usepackage{color}
\usepackage{graphics}
\usepackage{epsfig}
\usepackage[normalem]{ulem}
\usepackage{cancel}
\usepackage{float}

\usepackage[outercaption]{sidecap} 

\begin{document}
\title{Observation  of  1-D time dependent non propagating  laser plasma structures using  Fluid and PIC codes  }
\author{Deepa Verma, Ratan Kumar Bera, Atul Kumar, Bhavesh Patel and Amita Das}
\address{ Institute for Plasma Research, HBNI, Bhat, Gandhinagar-382428, India}

\begin{abstract}
The manuscript  reports the observation of  time 
dependent  localized and non-propagating structures in the coupled laser plasma system through 
 1-D fluid and PIC simulations. It is reported that such structures  form spontaneously as a result of collision amongst 
 certain exact solitonic solutions.   They  are seen to survive 
 as  coherent  entities for a long time ~ up to several hundreds of plasma 
 periods.   Furthermore, it is shown that such time dependence 
   can also be artificially recreated by significantly disturbing the delicate balance between the radiation and the density fields required for the exact non-propagating solution obtained by Esirkepov et al. \citep{bulanov_98}.  
 The ensuing time evolution is an interesting interplay between kinetic and field energies of the system. The electrostatic plasma oscillations are coupled with oscillations in the electromagnetic field. The inhomogeneity of the background and the relativistic nature, however, invariably produce large amplitude density perturbations leading to its wave breaking. In the fluid simulations, the signature of wave breaking can be discerned by a drop in the total energy which evidently gets lost to the grid. The PIC simulations are observed to closely follow the fluid simulations till the point of wave breaking.  However, the total energy in the case of PIC simulations is seen to remain conserved all throughout the simulations. 
At the wave breaking the particles are observed to acquire thermal kinetic energy in the case of PIC. Interestingly, 
 even after wave breaking,  compact coherent structures with trapped radiation inside high-density peaks,  continue to exist 
 both in PIC and fluid simulations. Though the time evolution does not exactly match in the two simulations as it does prior to the 
 process of wave breaking, the time-dependent features exhibited by the remnant structures are characteristically similar. 
\end{abstract}
\maketitle
\section{Introduction}
  In the last few decades the study of interaction between ultra intense laser pulse and plasma has attracted significant 
 amount of interest because of their applications in number of areas like laser fusion \cite{Tabak_94,mori_14}, 
 plasma-based particle accelerators \cite{liu_13,joshi_07,Santala_01,Tajima_79}  and photon acceleration 
 schemes \cite{Santala_2000} etc. Varieties of
 exact nonlinear  localized solutions have been observed in the extensive study of the laser plasma interaction 
 process \cite{Tabak_94,kaw_92, Farina_05,Mori_88, Katsouleas_88, Kuehl_93, sudan_97}.
  When a laser pulse interacts with the plasma medium, it expels plasma electrons due to ponderomotive force and creates a cavity evacuating plasma electrons from the center of a laser pulse. These electrons are pulled back by the stationary ions which are left in the background. Hence the exact balance between ponderomotive force and electrostatic force leads to a 
  configuration where the electrons are piled up at the edge and prevents the leak of radiation. 
Such a structure can either be stationary or be propagating with constant speed in the plasma. 
The complete characterization of these structures has been made earlier \cite{saxena_06, Poornakala_02, Farina_01, sita_11, deepa_15}. 
Depending on the number of peaks of the light pulse trapped in the density cage, these solutions have been termed as a single peak, paired and multiple peak solutions. The question of accessibility and stability of such structures has been addressed by fluid simulations \cite{Poornakala_02,saxena_06, saxena_07, sita_11}. Some PIC studies have also been carried out \cite{Farina_01,bulanov_98} on the evolution of coherent structures involving light matter coupled systems. 
The coherent structures are important in many ways.   
For instance, they can be used as a means for transporting energy in plasma medium.  
 They can also be utilized for particle and photon acceleration purposes \cite{jung_2011}. 
A precarious spatial balance of electrostatic and electromagnetic fields is required for the formation of exact solitonic structures. The frequency of the electromagnetic field and the propagation speed also have to satisfy stringent eigenvalue condition, for multiple peaks, only discrete eigenvalues are permitted. These conditions are very difficult to satisfy in a realistic situation.  It is shown that there also exist time-dependent localized structures which do not require any such delicate balance between the various fields. Such structures though time dependent,  are observed to survive as a single entity for a very long duration, e.g.  for several hundreds of plasma periods.  
The energy leakage is minimal. We feel that such versatile long-lived structures are also well suited for many applications including the transport of energy. 
The context of the formation of such time-dependent structures, the behavior they exhibit during evolution etc.,  have been examined in considerable detail using fluid and PIC simulations. In this manuscript, however,  we will restrict to non-propagating time-dependent structures. Propagating time-dependent structures have also been observed, the details of which will be presented in a subsequent manuscript. 
This paper has been organized as follows. The next section contains a  description of the governing equation along with a  brief discussion on numerical simulation and associated diagnostics.  
In section III, we provide a description of the spontaneously formed time-dependent non-propagating structure. We show that in the collisional interaction between high but unequal amplitude single peak structures,  the remnant structure displays interesting oscillatory behavior. These remnants 
are observed to invariably have a   high amplitude 
 (exceeding the upper limit of  radiation amplitude for static single peak solutions \cite{bulanov_98})  static oscillating profile. 
   In section IV, we recreate the nature of time dependence observed in such spontaneously formed structures by deliberately disturbing the delicate balance between the radiation and the electron density profile of the exact non-propagating solution proposed by Esirkepov et al \cite{bulanov_98}. 
     Despite significant disturbance added to the solution, the structure does not disintegrate but exhibits similar traits of time dependence as observed in a spontaneously formed structure in the aftermath of collisions between two exact solutions in section III. 
   A detailed study of these system shows that the energy alternates between field and kinetic forms. Basically, the excess radiation introduced in the system tries to leak out of the structure and in the process excites electron density oscillations. The excitation of plasma wave in an inhomogeneous density background maintained by the equilibrium structure along with relativistic effect leads to wave breaking. Both fluid and PIC simulations exactly match with each other before the onset of wave breaking. 
  In the event of wave breaking as the peaked density structures get formed the total energy evaluated from fluid description shows a small dip, whereas in the  PIC simulations the 
    energy continues to remain conserved. It is observed that around this time in PIC, the particles acquire random kinetic energy which is accounted for in the energy evaluation in PIC. 
    Interestingly even after wave breaking the radiation structure continues to retain its identity and the out of phase oscillations between field and kinetic energy continues. After wave breaking even though there is no exact match between the fluid and PIC time evolution, the characteristic features of the oscillations in terms of frequency etc., remains identical. 
Section V contains the summary and discussion. 
\section{Simulation set up}
\subsection{Fluid simulation}
The basic governing equations to study the formation and propagation of solitons in the coupled laser plasma system are the relativistic fluid-Maxwell equations. These equations contain the continuity equation, momentum equation for electrons and the full set of Maxwell's equations.
Here the dynamics of ions are ignored because of their slow response due to heavy mass. The intensity of the light
field (laser) is considered to be high enough for the electrons to get driven in the relativistic domain.
 This yields the following complete set of the equation in the normalized variables for variations restricted in 1-D, along with  $\hat{x}$ directions. 
\begin{equation}
   \frac{\partial n}{\partial t} +  \frac{\partial (nv_x) }{\partial x}= 0
\label{eq1}
\end{equation}

\begin{equation}
\left( \frac{\partial}{\partial t} + v_x\frac{\partial }{\partial x}\right)
(\gamma \vec v) = -\vec{E} - (\vec{v} \times \vec{B})
\label{eq2}
\end{equation}

 \begin{equation}
 \frac {\partial{\vec B}} {\partial t}=-\frac {\partial} {\partial x}(\widehat{x}  \times \vec{E}) 
\label{eq3}
 \end{equation}
 
 \begin{equation}
 \frac {\partial{\vec E}} {\partial t}= n \vec{v}+\frac {\partial} {\partial x}(\widehat{x}  \times \vec{B}) 
\label{eq4}
\end{equation}
 \begin{equation}
  \frac {\partial{E_x}} {\partial x}= (1- n)
 \label{eq5}
 \end{equation}
\begin{equation}
 \frac {\partial{B_x}} {\partial x}= 0
\label{eq6}
\end{equation}
where $\gamma=(1-v^2)^{-1/2}$, is the relativistic factor associated with the plasma electron of
density $n$ and velocity $\vec{v}$.  In above equations, $v_x$,  $E_x$ and $B_x$ represents the 
$x$- component of velocity $\vec{v}$, electric field $\vec{E}$ and magnetic field $\vec{B}$ 
respectively. We can also write,
$ \vec{E}=- \widehat{x} \frac {\partial \phi } {\partial x}-\frac{\partial \vec{A}}{\partial t}$ , 
$\frac {\partial^2 \phi } {\partial x^2}= (n-1)$ 
and  $\vec{B}=\frac {\partial} {\partial x}(\widehat{x}  \times \vec{A}) $, where $\phi$ and $\vec{A}$ 
represents electrostatic potential
and vector potential respectively. Here normalized variables  are, $n \rightarrow \frac{n}{n_0}$, 
$\vec{v}\rightarrow \frac{\vec{v}}{c}$, $\vec{E}\rightarrow \frac{e\vec{E}}{mc\omega_{pe}}$,
$\vec{B}\rightarrow \frac{e\vec{B}}{mc\omega_{pe}}$, $x\rightarrow \frac{x}{\frac{c}{\omega_{pe}}}$,
 $t\rightarrow {t \omega_{pe}}$, $\phi \rightarrow \frac{e\phi}{mc^2}$, $A \rightarrow \frac{eA}{mc^2}$.
The equations (\ref{eq1}-\ref{eq6} ) form the complete set of equations to study the light plasma interaction system in 1-D. 
A  fluid code was developed which is based on flux-corrected transport scheme 
\cite{boris_93} using the LCPFCT suite of subroutines. 
The basic principle of LCPFCT routine is based on a generalization of two-step Lax-Wendroff method \cite{Press_92}. 

\par\vspace{\baselineskip}

\subsection{Particle - In - Cell (PIC) Simulations} 
 The Particle -In - Cell (PIC) essentially uses the equation of motion (Lorentz force equation) for the evolution of particle velocity and position. The 
  Maxwell's equations are used to evolve electric and magnetic fields in a self-consistent manner.  The methodology of the PIC simulations has been described in detail in many review articles \cite{Dawson_83} and books \cite{Birdsall_85}. 
   
The box length of the system ($L_x$),  cell size ($\Delta x$) and time step ($\Delta t$) are chosen  similar for both fluid and 
PIC simulations.  The time step has been calculated using the Courant-Friedrich-Levy condition\cite{Courant_1928}. We have ignored the evolution of ions here. They only provide merely a smooth neutralizing background. The particle positions initially are chosen in such a fashion as to define the requisite electron density of the structure of choice. There are well-known prescriptions for the same \cite{Hockney_81}. The charge density and the current density defined at the grids are used for evolving the electric and magnetic fields. The electric and magnetic fields, in turn, are interpolated at the locations of the particle for advancing the velocity and the location of the particles through well-known schemes \cite{Birdsall_85}. 
The results from PIC simulations are in general quite noisy. A choice of about  $50$ to a maximum of about $250 $ particles per cell in our simulations showed a considerable reduction in noise. 
We present the results in terms of same normalizations as discussed in the fluid subsection. The boundary condition of the system has been taken to be periodic to perform this simulation.

We have initiated our simulation using the profiles of density, velocity, electric field and magnetic field for various kind of localized structures composed of coupled light and plasma medium. 
The profiles of density, velocity, electric field and magnetic field with time have been recorded.

\subsection{Energy and other diagnostics}
The total energy of the system is a crucial diagnostics for any simulation as it ensures the 
accuracy of the simulation. It can be shown that for the set of Eqs(\ref{eq1}-\ref{eq6}) the 
total energy composed of the sum of field  and kinetic energy of the particles satisfies the following 
conservation law: 
\begin{equation} 
\frac{1}{2}\frac{\partial}{\partial t} \int \left[ E^2 +B^2 \right] dx + \frac{\partial}{\partial t}\int \left[n(\gamma -1) 
\right] dx + \int \left[ \nabla \cdot (\vec{E} \times \vec{B}) \right] dx = 0
\end{equation}
The first term here is the field energy, the second term denotes the kinetic energy of the plasma 
medium. Here $\gamma$ is the relativistic factor. The third term represents the energy loss through Poynting flux. In our 
simulations, we have considered a sufficiently large box size to avoid the leaking of the radiation 
from the boundaries in the time scale of interest. Thus the sum of field and kinetic energy of the 
system should remain constant.

In the fluid simulations the  energies 
are evaluated numerically by summing up at the grid locations.  
$$ FE(t)= \frac{1}{2} \sum_i [E_i^2(t) +B_i^2(t)] \Delta x$$
$$KE(t)= \sum_i n_i(t) [\gamma_i(t) -1] \Delta x$$
$$TE(t)=KE(t) +FE(t)$$
here $i=1,2,3,...N_x$ and $N_x$ is the number of spatial grid points and  
$\Delta x $ denotes the grid spacing. Also,  
$E_i(t)$, $B_i(t)$, $n_i(t)$ and $\gamma_i(t)$ are the values of electric 
field, magnetic field, electron density and relativistic factor respectively at the $i-$th grid point 
at any time $t$. 

In the PIC simulations, the field energy is calculated in the similar fashion by summing the values
at the grid locations. However, for the kinetic energy,  the sum is over all the particles. Thus, we have 
\begin{equation}
KE(t)=\sum_j \gamma_j(n_j-1) 
\end{equation}
Where $j=1,2,...N_p$ represents the index associated with the electrons in the system. 
The change in kinetic is related to the work done by the electric field on particles  (electrons here). 
We thus have 
\begin{equation}
\frac{\partial}{\partial t}\int \left[n(\gamma -1) \right] dx = - \int n \vec{v} \cdot \vec{E} dx = - \frac{1}{2}\frac{\partial}{\partial t} \int \left[ E^2 +B^2 \right] dx 
\end{equation}

\section{Spontaneously excited  non-propagating time-dependent structures} 
The coupled set of laser plasma equations permits a variety of stationary and propagating exact
solutions which have been characterized in
the parameter space of frequency vs. group velocity space \cite{kaw_92,saxena_06}. 
As mentioned earlier, these solutions are interesting and tell us that it is possible in some cases for the radiation to move in a plasma such that the plasma wakefield excited at its front gets absorbed by its tail. For structures with zero group speed, however, the front and tail cannot be distinguished and the light wave here is simply trapped in an electron density cavity that it digs for itself by ponderomotive pressure.  Exact analytical form of such solutions have been obtained by 
 Esirkepov et al \cite{bulanov_98}. They showed that there is an upper limit on the amplitude of such structures which define complete cavitation. 

\begin{figure}[H]
   \vspace{3cm}
\includegraphics[height=10.0cm,width=19.0cm]{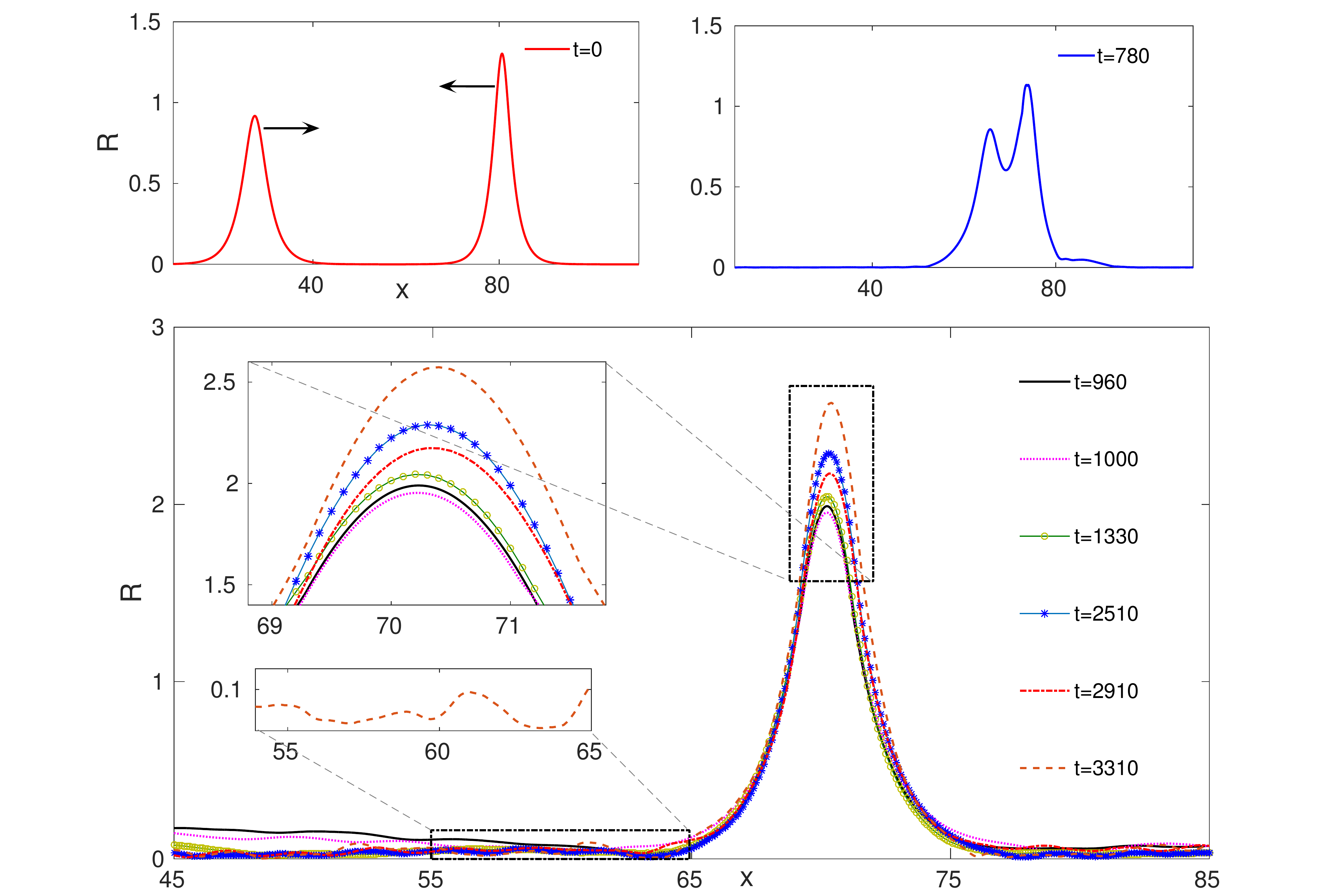}
\caption{Collision of two single peak of soliton moving at group velocity $\beta=-0.01$ and $\beta=0.05$.}
\label{figure_1}
\end{figure}

The evolution and collisional interaction of some of these structures have been studied \cite{saxena_06,saxena_07}. At low amplitudes, the behavior of these structures is similar to solitons. 
At high amplitudes, there is a perceptible difference in their behavior. In fact, we show here that when two high but unequal amplitude oppositely
moving exact single peak structures suffer collision as shown in Fig.(\ref{figure_1}), they merge together and form a time dependent non-propagating structure. 
The amplitude of this spontaneously formed structure shows oscillations in time and there is also evidence of the radiation leaking out from it. 
However, the overall structure seems to persist for a long time. 

The amplitude of radiation, in this case,  exceeds the upper permissible limit of static exact solutions [ Esirkepov et al., \cite{bulanov_98}] and triggers
a complex interplay of plasma density and radiation field oscillations. It should be noted that even though the original structures had
group speeds in opposite directions of unequal magnitude the final structure is non-propagating. This has been verified in a number of such collisional studies. It appears that there is a preference
towards formation of non-propagating structures. These non-propagating time-dependent structures survive for a long time. 
This seems to suggest that the time-dependent structures are more realistic. They form readily, do not require a 
precarious balance of various fields and survive for a considerable time.

In the next section, we specifically choose the non-propagating exact structures and deliberately introduce a perturbation in it ( in terms of an enhanced radiation) to observe its subsequent evolution. 


\section{Time dependence  of  Esirkepov structures with enhanced radiation}
Esirkepov et al. \cite{bulanov_98} have obtained an exact analytical form for the stationary solitonic structures. We take the analytical form of the solution and express it in terms of the initial conditions for electric, magnetic and velocity fields.  These fields are then evolved in our simulations. 
The initial density of the electron is chosen to satisfy the analytical form of the solution. 
The  stationary solitary solutions in our simulation was chosen for a value of $A_0=1$;  
where $A_0$ is the peak value of $R$ of the solution. The frequency of laser has been derived from, 
$\omega=\sqrt{2\sqrt{1+A_0^2}-2}/A_0 \simeq 0.9102$.  We have verified the stationarity and energy conservation of
these exact solutions with both our fluid and PIC codes. 

\begin{figure}[H]
   \vspace{1cm}
\includegraphics[height=7.0cm,width=19.0cm]{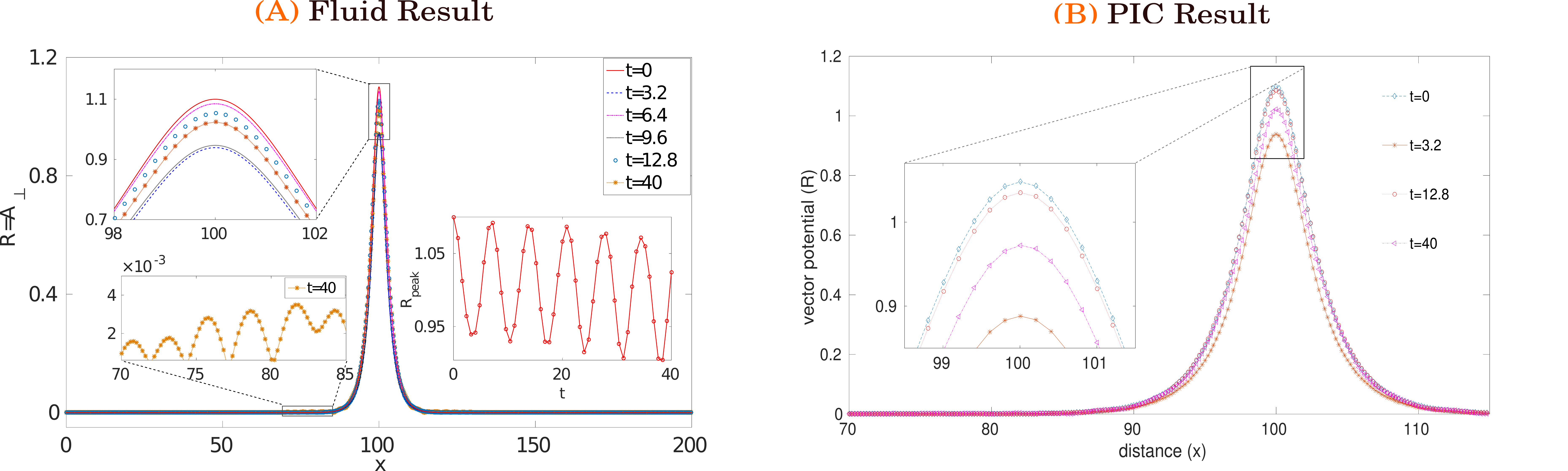}
\caption{Vector potential ($R$) for stationary ($\beta$=0) single peak soliton of amplitude $R=1.1R_0$, confine an excess radiation $10\%$ of $R_0$ in electron density cavity at different times. }  
\label{figure_2}
\end{figure}

\begin{figure}[!hbt]
\includegraphics[height=7.0cm,width=19.0cm]{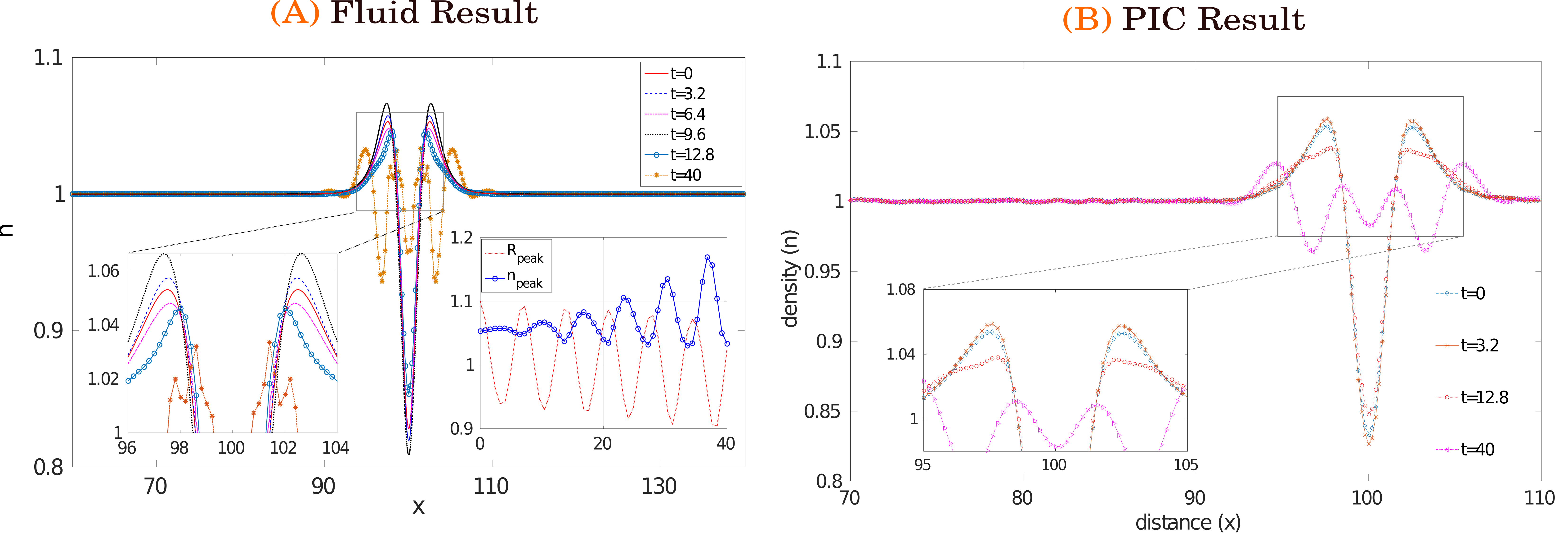}
\caption{Plot of density ($n$) for stationary ($\beta$=0) single peak soliton associated with an excess radiation of amount $10\%$ of $R_0$ at different times. } 
\label{figure_3}
\end{figure}

We now take the same solution, keep all the fields identical but enhance the trapped radiation inside by a multiplicative factor of $1.1$. This disturbs the precarious
parallel force balance of the electron momentum equation leading to oscillations in electron density. 
The evolving profiles for  $R$ and $n$ are shown in figure~(\ref{figure_2}) and figure~(\ref{figure_3}) respectively from both fluid and PIC simulations. 
The two simulations are in remarkable agreement. The oscillations in density and radiation field $R$ seem to be quite regular. 
In one of the subplots of 
 figure~(\ref{figure_3}), we have shown the evolution of the peak of both density $n$ and $R$ fields. It should be noted that the oscillations of the two fields are typically always out of phase in time. It can be understood by realizing that the excess radiation trapped inside the structure offers excess ponderomotive pressure due to which the electron density is pushed out and starts getting enhanced at the edge.  As the density cavity gets deeper and broader the radiation trapped at the center expands out leading to the fall in its peak. This leads to a drop in ponderomotive pressure which is now dominated by the electrostatic pull acting on electrons by the ions. Thus the plasma oscillations triggered by the excess radiation continue in time. 
 It should be noted that these plasma oscillations are occurring against an inhomogeneous background density profile sustained by the trapped radiation. 
 Moreover,  the relativistic $\gamma$ factor is also a function of space. Thus such an oscillation would be expected to suffer wave breaking. 

\begin{figure}[H]
 \vspace{3cm}
\includegraphics[height=10.0cm,width=19.0cm]{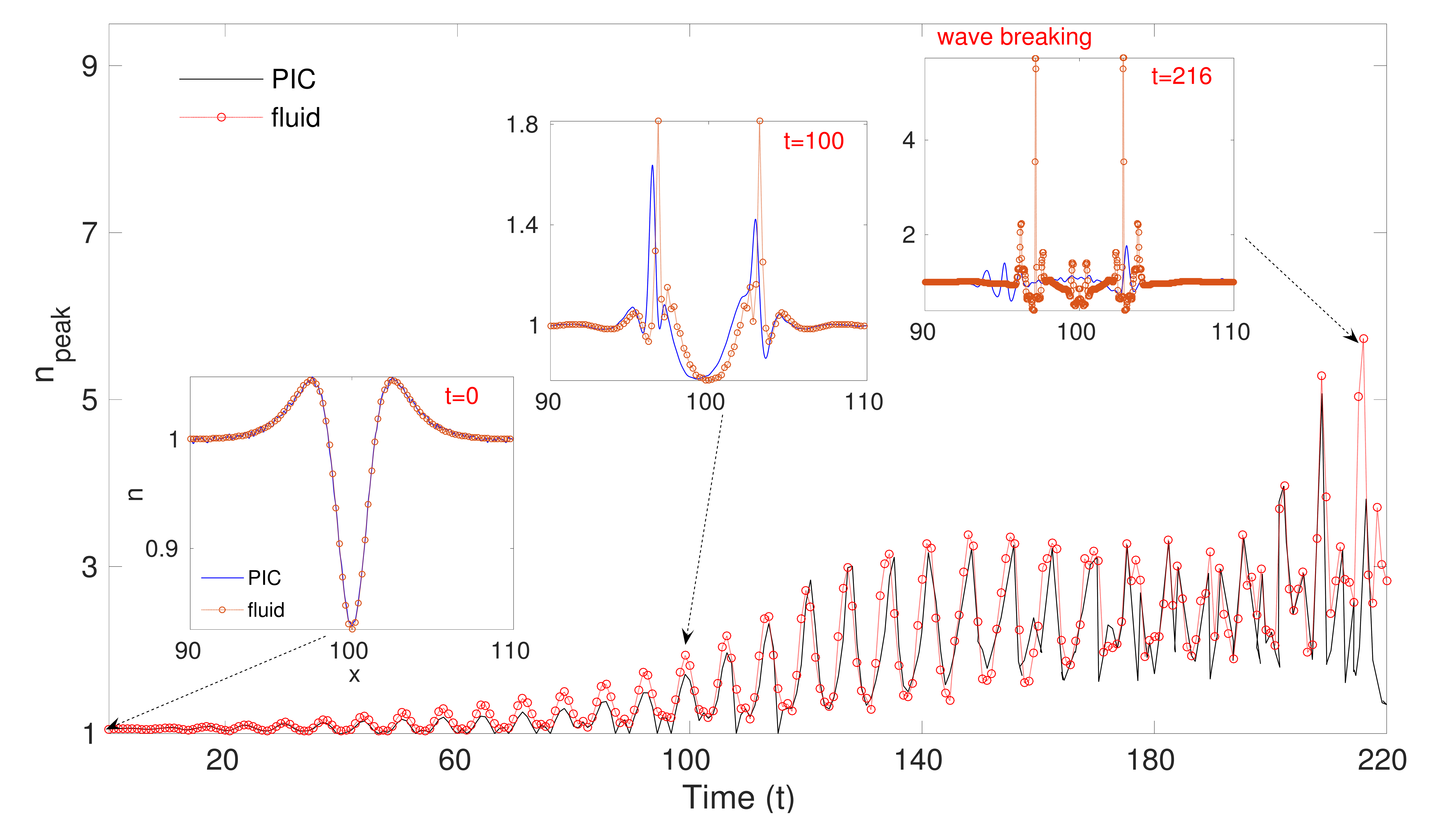}
\caption{density peak $n_{peak}$ oscillation as a function of time $t$ for the confinement of a symmetric excess radiation (of amount $10\% $) of stable vector potential in electron cavity. } 
\label{figure_4}
\end{figure}

\begin{figure}[H]
   \vspace{3cm}
\includegraphics[height=10.0cm,width=19.0cm]{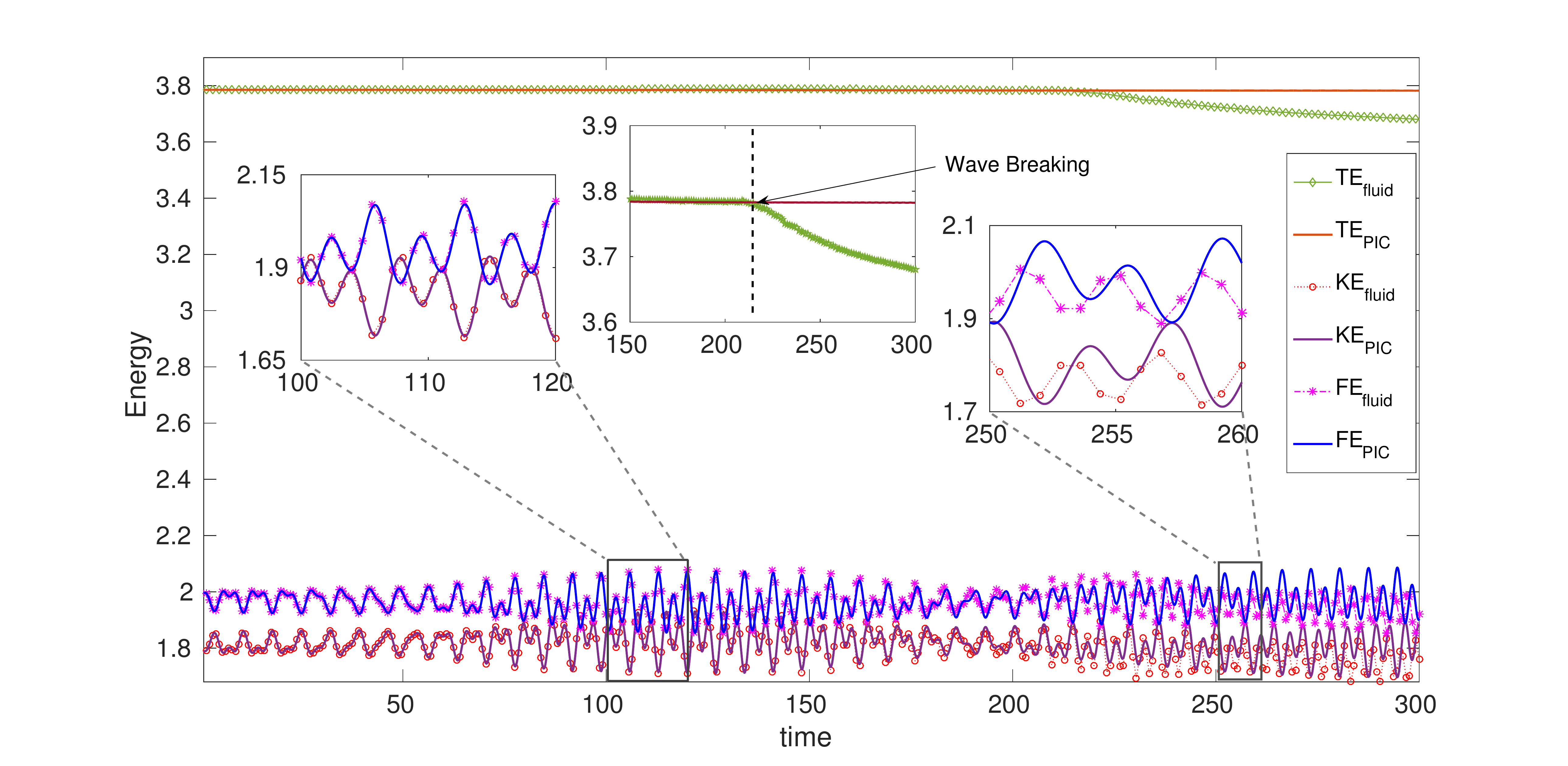}
\caption{Kinetic energy (KE), Field energy (FE) and total energy (TE) for the confinement of a symmetric excess radiation (of amount $10\% $) of stable vector potential in electron cavity as a function of time  $t$ from fluid(marker with dashed) and PIC(solid line). } 
\label{figure_5}
\end{figure}

 From the time dependence of the peak of the radiation field  $R$,  it can be observed that it steadily decreases. This happens as a result of  steady  leakage of the radiation from the 
 edges (see 
 inset of figure~(\ref{figure_2}) where the edge portion of the structure has been zoomed in at $t = 40$) 
 indicating clearly the leaking of the radiation. The peak of density oscillations, however,  steadily increases with time. 
 In fact, the density profile is observed to generate several sharply peaked structures.  
   Around $ t\sim 216$ as shown in figure~(\ref{figure_4}), the density acquires a very spiky form. 
This, in fact, is a signature of wave breaking phenomena. We have tracked the   
  total energy (TE) evolution in Fig.(\ref{figure_5}) which is conserved all throughout but shows a 
  very small dip exactly at $ t\sim 216$ in the fluid simulation, exactly at the same time when the spike in density spike gets formed. 
  Despite changing the grid size in the fluid simulation the energy dip and the density spike typically occurs around the same time.

It is interesting to observe that in the PIC  simulations too exactly around this time the density spikes appear. 
  At the wave breaking point in the fluid code, the energy gets lost to the grid. 
  In PIC simulations where the total energy incorporates the individual kinetic energies of the particles,  the total energy remains conserved (See figure~(\ref{figure_5})). 
  However, this energy now shows up as the random kinetic energy of the particles.
    It is interesting to note that the FE and KE continue to remain out of phase before as well as after and also during the wave breaking process as can be viewed from the enhanced inset of the Fig.(\ref{figure_5}). The evolution of FE and KE match exactly in the fluid and PIC simulations before wave breaking. However after wave breaking there appears a slight mismatch between the fluid and PIC simulations. 
  
\begin{figure}[H]
   \vspace{1cm}
\includegraphics[height=7.0cm,width=19.0cm]{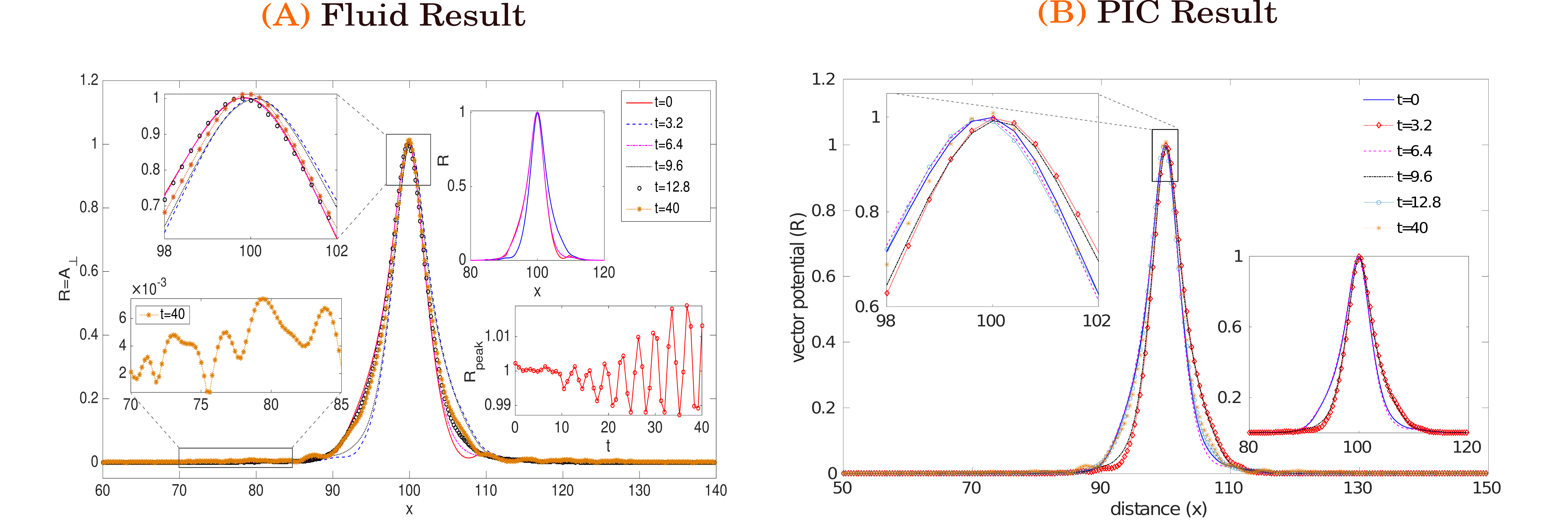}
\caption{Vector potential (R) profile having an asymmetricity of $10\%$ in the laser
pulse from its stable solution. }  
\label{figure_7}
\end{figure}

  We also considered perturbing the radiation field asymmetrically in the structure. 
 This choice ensures that at one of the edges radiation pressure dominates whereas at the other edge scalar potential is the dominant force. In this case, from Fig~(\ref{figure_7}),  
 one observes that the structures show asymmetric oscillations with one edge expanding while the other contracts. 
Basically, the edge where the radiation pressure exceeds the equilibrium value, the radiation tends to expand out. 
At the other edge where the $R$  is lower than
equilibrium value, it is pushed in. Some amount of radiation is again observed to leak out. The asymmetric oscillations
are in much closer qualitative agreement with the results reported in
section III where the collision between unequal structures led to the formation of non-propagating time-dependent structure.

\begin{figure}[H]
   \vspace{1cm}
\includegraphics[height=10.0cm,width=19.0cm]{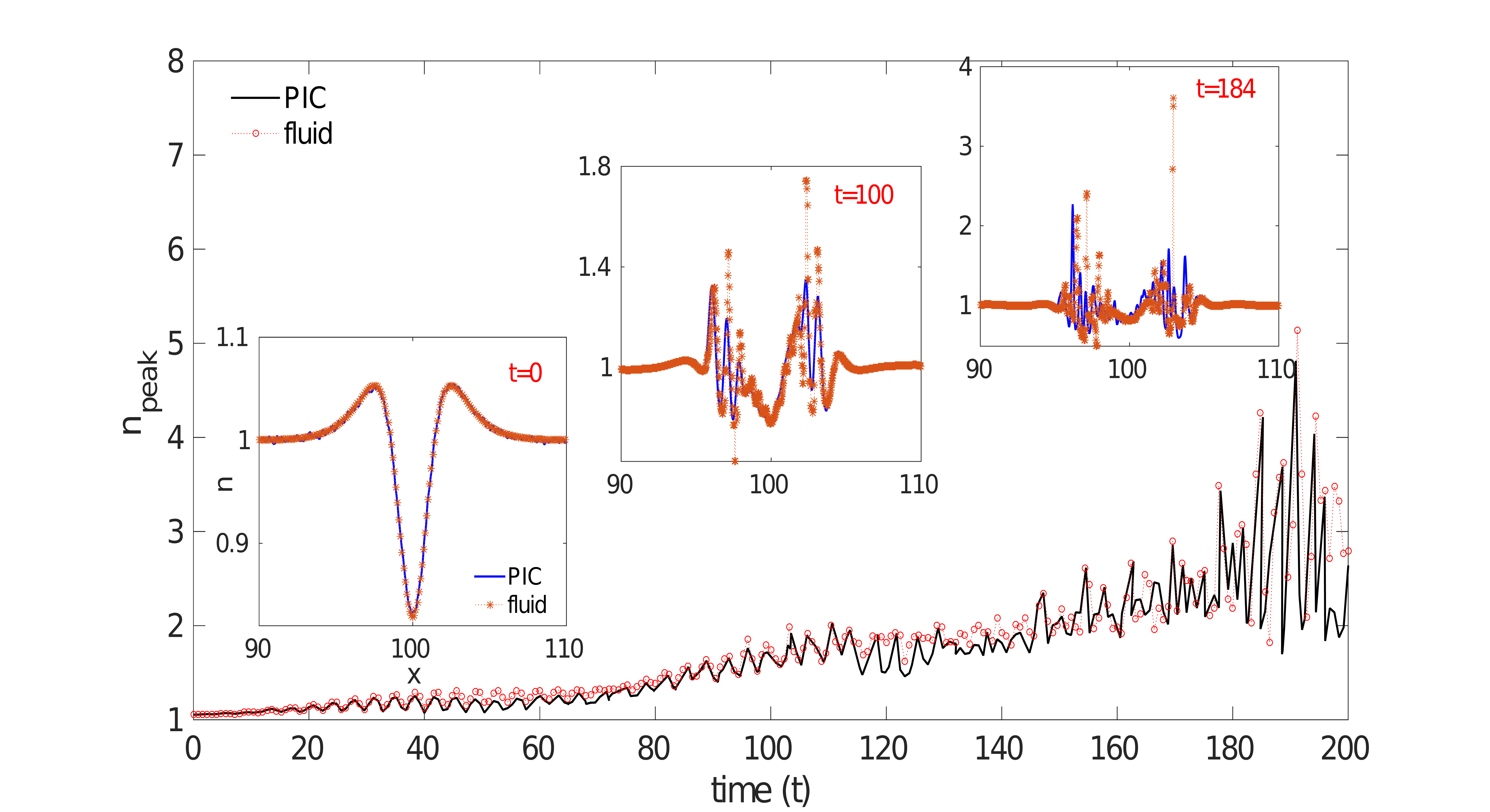}
\caption{density peak $n_{peak}$ oscillation as a function of time $t$ for the confinement of a assymmetric excess radiation (of amount $10\% $) of stable vector potential in electron cavity. }  
\label{figure_8}
\end{figure}

\begin{figure}[H]
   \vspace{1cm}
\includegraphics[height=10.0cm,width=19.0cm]{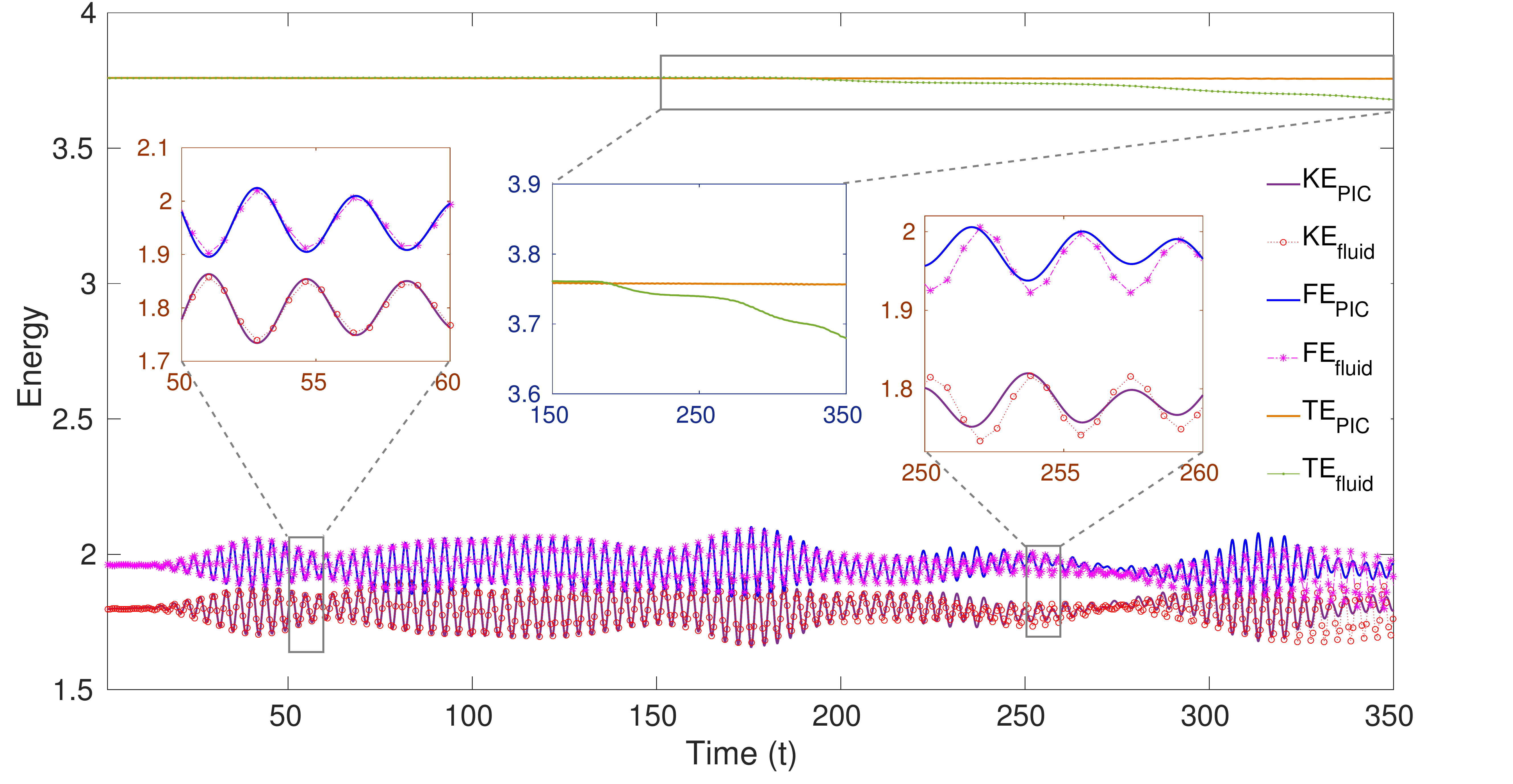}
\caption{Kinetic energy (KE), Field energy (FE) and total energy (TE) for the confinement of a assymmetric excess radiation (of amount $10\% $) of stable vector potential in electron cavity as a function of time  $t$ from fluid(marker with dashed) and PIC(solid line). }  
\label{figure_9}
\end{figure}

The plasma oscillations also get excited and the amplitude of density keeps growing. Ultimately wave breaking occurs at about $t \sim 184$ which is tracked by the formation of
density spikes (see fig~(\ref{figure_8})) and a dip in the value of the total energy which can be seen from figure~(\ref{figure_9}). 
Thereafter one again ends up with structures in which radiation trapped between density peaks survives for a long duration.

\section{Summary and Discussion}
We report  observations of time-dependent  localized structures in the 
1-D fluid as well as PIC simulations for the coupled laser plasma system.  
 Despite such a time dependence the structures are found to be fairly robust in a sense that they survive by
retaining their identity for several hundreds of plasma periods. 
Such time-dependent structures can form either spontaneously or can also be recreated by disturbing the delicate balance between various fields required in the context of exact solutions. 
For instance,  the collision amidst two high but different amplitude exact solutions also leads to the formation of a non-propagating structure with oscillating amplitudes. 
the same is also observed when the exact non-propagating solutions obtained by Esirkepov et al \cite{bulanov_98}  are deliberately disturbed by introducing excess radiation. 

It should be noted that while the exact solutions require a very delicate balance between the radiation and density fields for the time-dependent structures, it
is not necessary to satisfy such a stringent condition. Thus while it would be rather difficult to form exact solutions experimentally,  in contrast, these time-dependent structures which form spontaneously and retain their identity for a long time would be more easily amenable in experimental observations. 

This raises several questions. Does the laser plasma system permit a new variety of time-dependent solutions, where the time dependence is not merely restricted to steady propagation? 
We also have evidence of obtaining time dependent propagating structures spontaneously which will be reported in a subsequent publication.

 

\end{document}